\documentstyle[spie,times]{article}
\input{psfig}

\title{Interoperability tools for the Virtual Observatory}

\author{Daniel Egret\supit{a} and Fran\c{c}oise Genova\supit{a}
\skiplinehalf
\supit{a}CDS, UMR 7550, Observatoire de Strasbourg,
11 rue de l'Universit\'e, 67000 Strasbourg, France}

\authorinfo{E-mail: Egret@astro.u-strasbg.fr, Genova@astro.u-strasbg.fr}

\begin{document}
\maketitle

\begin{abstract}
Interoperability is one of the key 
issues in the current efforts to build the Virtual Observatory.
We present here some of the tools which already contribute
to the efficient exchange of information between archives,
databases, and journals.
\end{abstract}


\section{Introduction}

Interoperability has been identified as one of the crucial 
issues of the emerging Virtual Observatory
at an international level
(for an introduction to the concept of the
Virtual Observatory, see Brunner et al. 2001\cite{caltech}).

The challenge is to provide the scientific community 
with an integrated access   to a wide variety of
information, databases and services, not only heterogeneous 
by their origins
(observatory archives, electronic publications,
national or international services) 
but also diverse
in nature: bibliography, images, data tables, spectra, etc.

Solutions for interoperability imply a strong coordination of
all data producers and archive managers, so that common protocols
and standards can be discussed, decided, and implemented.

At a basic level, interoperability requires the ability
for a database system to \emph{understand a query} formulated
by another complementary database, to \emph{process it
unambiguously}, and to \emph{transfer the results} back 
to the querying system so that 
it can be used for building an 
integrated response to a user.

Let us give some examples of these different steps:

\begin{itemize}
\item   \emph{to understand the query}: this implies that the query
syntax is correct. 
Most of the past efforts (see e.g., Astrobrowse, section~\ref{astrobrowse})
have been devoted to the management of queries by position
because most of the databases
share an approach by equatorial coordinates.
The ASU protocol (Ochsenbein et al. 1998\cite{asu}) described 
in section~\ref{sec:asu}
is an example of a coordinated effort to standardize query formulation
for submission to Web servers.

\item \emph{to process a query unambiguously} it is important
to ensure that potentially interoperable systems 
share the same assumptions (e.g. : equatorial coordinates
are expressed in the J2000 system).
At a first level, this can start with very simple
agreed upon data description; a striking example is the bibcode:
this 19-character code (Schmitz et al.\ 1995\cite{bibcode}) 
describes unambiguously a reference
to a paper published in a journal or book within the astronomy
and astrophysics scientific community.
As a second step, it becomes essential that interoperable databases
and services share common \emph{data dictionary} parameters;  
this is the main driver
for the GLU project which will be described in section~\ref{glu}.
For instance, the GLU system will keep the knowledge
about the coordinate system used by each service, and will
provide the coordinate translation from the query, when appropriate.

\item \emph{transfer the results for further use} implies 
in principle a symbolic --as well as a physical-- description of all retrieved
data items.  In practice most systems simply deliver a
status code (or a number of records) and the corresponding
records (or an access to these records on a remote site).

It is part of the future
challenge of the Virtual Observatory to be able
to go much further in this direction, and reach a point
where all data elements would be rigorously described
so that they can be used within classification
algorithms, for example, as a step of knowledge extraction.

\end{itemize}

\section{Protocols and standards}

The first step towards interoperability is the development
of common protocols and standards for the astronomical
information services.
As shown, e.g. by Egret et al. (2000\cite{ehm}),
the history of coordinated development of astronomical 
information services demonstrates that
coordinating spirit is not out of reach in a small community such
as astronomy, largely sheltered from
commercial influence.

In this section we give some examples of this coordinated effort
which constitute some of the keys for the Virtual Observatory 
of the future: 
a standard format (FITS); 
a standard descriptor (bibcode); 
and a proposed syntax for web queries (ASU).

Finally we present AstroRes, a proposed data description for XML.

\subsection{FITS data and image format}

The Flexible Image Transport System 
(FITS) for astronomy (Wells et al.\ 1981\cite{fits})
is a commonly agreed system to encode both a definition of the data 
and the data itself in a machine independent way.
It is a clear and unambiguous standard for stating how geometric
information in an astronomical image should be represented.

The advantage of using a standard format for transport of 
astronomical images was soon realized and most major
observatories implemented it as the prime format for data exchange. 
Subsequently, the FITS format was recommended 
by International Astronomical Union as 
the standard format for interchange of image data between all observatories.

The extension of the FITS format to table and catalogue data 
resulted from the work of a Task Force appointed under IAU recommendation,
and is described by Grosb{\o}l et al.\ (1988\cite{grosbol})
and Harten et al.\ (1988\cite{harten}). 
The more recent adoption of the World Coordinate System (WCS) standard
(Calabretta and Greisen 2000\cite{wcs}) is a necessary
tool for all kinds of astrometric cross-matching of images and catalogues.

While FITS is `the' reference format for astronomical image description,
it is certainly less successful to describe data tables, 
where simpler robust systems are frequently preferred.

One of them is the {\tt ReadMe} file system implemented by Ochsenbein
(1994\cite{readme}) and used by the astronomical data centers around the world.
This file description is also used by the VizieR catalogue database, and for
the description of data tables associated to all major astronomical
journals.
Similarly to FITS, it uses a header file to describe the data,
but the main data file is simply coded in ASCII, without any
overhead. The header is easily understandable by the human, as well
as by the computer.

\subsection{Reference code}

Handling references to the literature is a very common, and
often confusing, task. The astronomers have adopted 
a very simple system, the \emph{bibcode}, to describe
unambiguously a reference
to a paper published in a journal or book within the astronomy
and astrophysics scientific community.
This 19-character code (Schmitz et al.\ 1995\cite{bibcode})
contains enough information (and redundancy) to allow an immediate
identification of a reference.

Example: {\tt 2000ApJS..131..335R} is the {\em bibcode} for a
paper\cite{Xid} published in the year 2000, in the {\em Astrophys. Journal Suppl.}
(ApJS), vol.\ 131, p.\ 335. The final letter ({\tt R}) is the initial
letter of the name of the first author (Rutledge).

The fact that major databases (NED, SIMBAD, ...), abstract service
(ADS), and electronic publishers (Journals of the American Astronomical 
Society, \emph{Astronomy \& Astrophysics}, ...),
all members of the \emph{Urania}\footnote{http://www.aas.org/Urania/}
collaboration,
share the same system has been a very
powerful help for a rapid development of interoperability
between abstract services, journals and databases.

\subsection{Astronomical Standardized URL}
\label{sec:asu}

The Astronomical Standardized URL 
(ASU\footnote{http://vizier.u-strasbg.fr/doc/asu.html},
Ochsenbein et al. 1998\cite{asu}) 
results from discussions 
between several institutes (CDS, ESO, ESA, CADC, OAT), 
for proposing a common syntax for on-line Web queries (generally
using the GET method for specifying query parameters directly
in the URL).

The basic concept of ASU is a standardized way of specifying 
query parameters such as:

\begin{itemize}
\item  catalogues: {\tt -source=catalog\_designation}, 
\item target positions: {\tt -c=name\_or\_position}, 
\item and radius: {\tt rm=radius\_in\_arcmin},
\item output format:  {\tt -mime=type},
\item and general constraints on parameters: {\tt column\_name=constraint}. 
\end{itemize}

Example: To get the X-rays sources from the Rosat All-Sky Bright 
Source Catalogue around M31, you can use the following URL:
http://vizier.u-strasbg.fr/cgi-bin/VizieR?-source=1RXS\&-c=M31,rm=20\&-out.all 

ASU is currently used within a number of archives and catalogue services,
generally in parallel with other, more sophisticated, or
more system-dependent query mechanisms.

\subsection{XML AstroRes}
\label{astrores}

The Extensible Markup Language (XML)
is a developing standard in which the description of the data (the
metadata) is included with the actual data in a single electronic document. 

XML is an ideal support for developing new standards 
for accessing and understanding tabular data, particularly 
for handling the responses from queries to on-line
catalogue services. 
If such responses are encoded in XML using agreed upon tags and attributes, 
it becomes possible to both display the data in clearly formatted tables and 
use the data in other applications (such as generating
graphical overlays of object positions on survey images). 

XML-encoded tables can also provide the basis for the
next generation of data discovery and integration tools (Astrobrowse, ISAIA,
see below). 
A certain number of
initiatives are under way to develop a general frame for XML in
astronomy (see, e.g., XML for Astronomy at 
NASA/GSFC\cite{xml-gsfc}
and XML resources at NASA/ADC\cite{xml-adc}).

Some developments tackle well defined
questions for implementing operational tools. 
For instance {\sc Aladin} (section~\ref{sec:aladin}) is fully XML compatible. 
The AstroRes standard, developed as the result of a
collaboration between scientists from CDS, ESO, Univ. Illinois,
STScI and GSFC provides a 
first standard data description for astronomical tabular outputs.
The detailed definitions of XML AstroRes
tags and attributes, including examples and the proposed
Data Description Document, are available 
on-line\footnote{http://vizier.u-strasbg.fr/doc/astrores.htx}
(Ochsenbein et al.\ 2000\cite{astrores}).

\section{Astronomical data dictionaries}

With a large
number of on-line services giving access to data or information, it is
clear that tools giving 
coordinated access to several (or even many) distributed services 
are needed. This
was, for instance, the concern expressed by NASA through the Astrobrowse
project (Heikkila et al.\ 1999\cite{astrobrowse-2}). 

In this section we will first describe a tool for managing
a ``metadata'' dictionary of astronomy information services
(GLU); then we will show how the existence of such a
metadatabase can be used for building efficient search
and discovery tools.

\subsection{The CDS GLU}
\label{glu}

The CDS (Centre de Donn\'ees astronomiques de Strasbourg) has 
developed 
GLU\footnote{http://simbad.u-strasbg.fr/glu/glu.htx}
(G\'en\'erateur de Liens 
Uniformes, i.e. Uniform Link Generator;
Fernique et al.\ 1998\cite{glu})
as a tool for managing remote links in a context of 
distributed heterogeneous services.

   \begin{figure}[htbp]
   \begin{center}
   \begin{tabular}{c}
   \psfig{figure=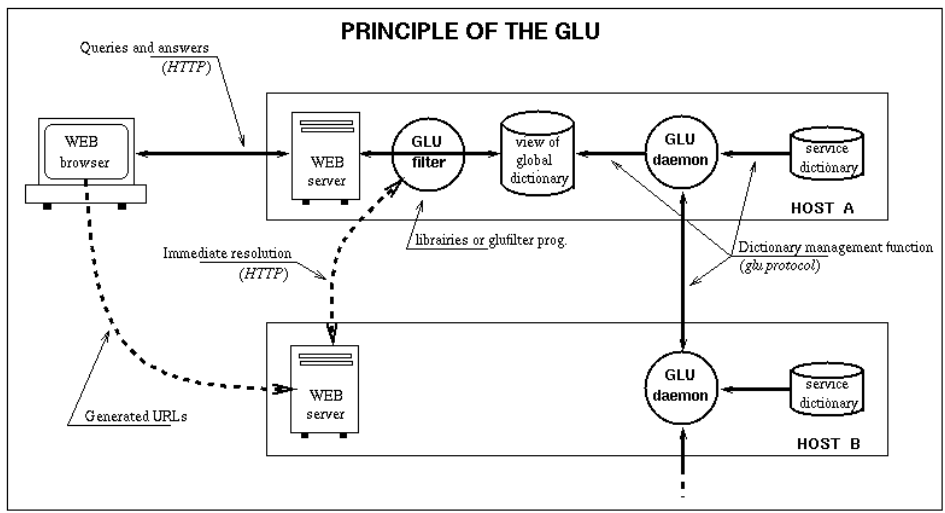,height=7cm}
   \end{tabular}
   \end{center}
   \caption[GLU]
   { \label{fig:glu}
A scheme of the GLU principles.}
   \end{figure}

First developed for
ensuring efficient interoperability  of the several services existing at CDS 
({\sc VizieR},  {\sc Simbad}, {\sc Aladin}, bibliography, etc.; see
Genova et al.\  2001\cite{cds}), 
this tool has also been designed for maintaining addresses (URLs) 
and query syntax,
and for storing a view of the conceptual data model
of distributed services 
(ADS, NED, observatory archives, journals, etc.).

A key element of the system is the ``GLU dictionary'' maintained by 
the data providers contributing to the system, and distributed to all 
sites of a given domain. This dictionary contains knowledge 
(meta-data) about the participating services 
(URLs, syntax and semantics of input fields, 
descriptions, etc.), so that it is possible to generate automatically a 
correct query for submission to a remote database. 

The service provider (data center, archive manager, or  
webmaster of an astronomical institute) can use GLU for 
coding a query, taking benefit of the easy update of the system: 
knowing which service to call, and which answer to expect from 
this service, the programmer does not have to worry about the 
precise address of the remote service at a given time, nor of the 
detailed syntax of the query (expected format of the equatorial 
coordinates, etc.).
The system includes a mechanism to share and maintain the distributed 
dictionary, similarly to what is done for name services on the Internet.

\subsection{New search and discovery tools}

The example of GLU demonstrates the usefulness of storing into
a database the knowledge about information
services (their address,  purpose, domain of coverage,
query syntax, etc.). In a second step, such a
database can be queried when the challenge is to provide
information about whom is providing what, 
for a given object, region of the sky, or domain of interest.

Several projects are working to provide prototype solutions,
based on the concept of a \emph{data access layer}
where general queries are translated into the 
multiple specific query languages
of individual services.
We will present in this section \emph{AstroBrowse} and \emph{AstroGlu}.
Other examples are {\sc Amase}\footnote{http://amase.gsfc.nasa.gov/}
at NASA/ADC, or \emph{Querator}, described by Pierfederici et al.\
in this conference.

\subsubsection{Astrobrowse}
\label{astrobrowse}

\emph{Astrobrowse} is a project that began within the United States
astrophysics community, primarily within NASA data centers, for developing
a user agent which significantly streamlines 
the process of locating astronomical data on the web. 
Several prototype implementations are already 
available\footnote{http://heasarc.gsfc.nasa.gov/ab/}.
With any of these prototypes, a user can already 
query thousands
of resources  without having to deal with out-of-date URLs, 
or spend time figuring out how to use each resource's
unique input formats. 
Given a user's selection of 
web-based astronomical databases and an 
object name or coordinates, Astrobrowse will 
send queries to all databases identified as containing potentially
relevant data.  It provides links to these resources and allows the
user to browse results from each query.  Astrobrowse does not recognize,
however, when a query yields a null result, nor does it integrate
query results into a common format to enable intercomparison.

\subsubsection{AstroGLU}

Consider the following scenario: we have a data
item $I$  (for example an author's name, the position or name
of an astronomical  object, a bibliographical reference, etc.),
and we would like to know  more about it, but we do not know a
priori which service $S$ to contact, and  what are the
different data types $D$ which can be  received in response.
This scenario is typical of a scientist
exploring new domains as part of a research procedure.

The GLU dictionary can actually be used for helping to solve this 
question: the dictionary can be considered as a reference 
directory, storing the knowledge about all services accepting data item $I$ as 
input, for retrieving data $D_1$ or $D_2$.  For example, we can 
easily obtain from such a dictionary the list of all services accepting 
an author's name as input; information which can be accessed, in 
return, may be an abstract (service ADS), a preprint (LANL/astro-ph), 
the author's address (RGO e-mail directory) or personal
Web  page (StarHeads), etc.

Based on such a system, it becomes possible to create automatically 
a simple interface guiding the user towards any of the services 
described in the dictionary.

This idea has been developed as a prototype discovery tool, under the name of 
AstroGLU\footnote{http://simbad.u-strasbg.fr/glu/cgi-bin/astroglu.pl}
(Egret et al.\ 1998\cite{astroglu}).
The aim of this tool is to help the users find their way among 
several dozens (for the moment) of possible actions or services.
A number of compromises have to be taken between providing the 
user with the full information (which would be too abundant and 
thus unusable), and preparing digest lists (which implies hiding some
amount of auxiliary information and making somewhat subjective
selections).

A resulting issue is the fact that the system puts on the same line 
services which have very different quantitative or qualitative 
characteristics.  {AstroGLU} has no
efficient ways yet to provide the user with a hierarchy of
services, as a gastronomic guide would do for restaurants.
Qualification of relevant datasets and services, 
on the basis of well-established science requirements,
will be a necessary step of the Virtual Observatory.

\section{Towards an integration of distributed data 
         and information services}

To go further, one needs
to be able to integrate the result of queries 
provided by heterogeneous services.

In this section we will first present
working examples of data integration --the
{\sc Simbad} or NED name resolvers, and the {\sc Aladin} 
interactive sky atlas.
Then we will describe the ISAIA project of an
integrated system able to display the results of complex
heterogeneous queries.

\subsection{Name resolvers}

A name resolver is a basic example of interoperability:
it is a simple program which sends the name of an astronomical
object to the SIMBAD or NED databases, and receives in return
the position of the object on the sky (or an error message 
when the name is not known from the database).
Such programs have been made available, even before the advent
of the web, as simple and robust client/server routines.

   \begin{figure}[htbp]
   \begin{center}
   \begin{tabular}{c}
   \psfig{figure=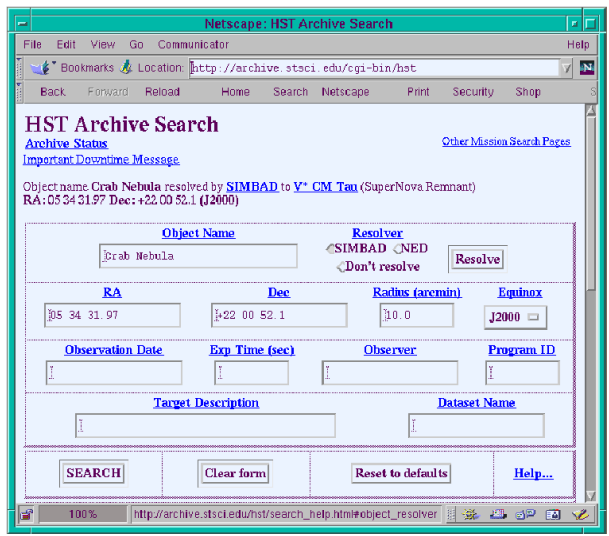,height=7cm}
   \end{tabular}
   \end{center}
   \caption[Resolver]
   { \label{resolver}
An example of SIMBAD/NED name resolver in the Hubble Spacec Telescope
archive system at STScI.}
   \end{figure}

NED and SIMBAD name resolvers are now used throughout the community,
integrated into most of the astronomical on-line archives and
databases (Fig.~\ref{resolver}).

\subsection{The example of ALADIN}
\label{sec:aladin}

The {\sc Aladin} interactive sky atlas
developed by the CDS in Strasbourg is a powerful prototype of data
integrators.

{\sc Aladin} has been primarily
developed for the identification
of astronomical sources through visual analysis
of reference sky images (Bonnarel et al.\ 2000a\cite{aladin}).
{\sc Aladin} fully benefits from the
environment of CDS databases and services
({\sc Simbad} reference database, VizieR catalogue service,
etc.), and is designed as a multi-purpose
service for use by the astronomical community worldwide.

   \begin{figure}[htbp]
   \begin{center}
   \begin{tabular}{c}
   \psfig{figure=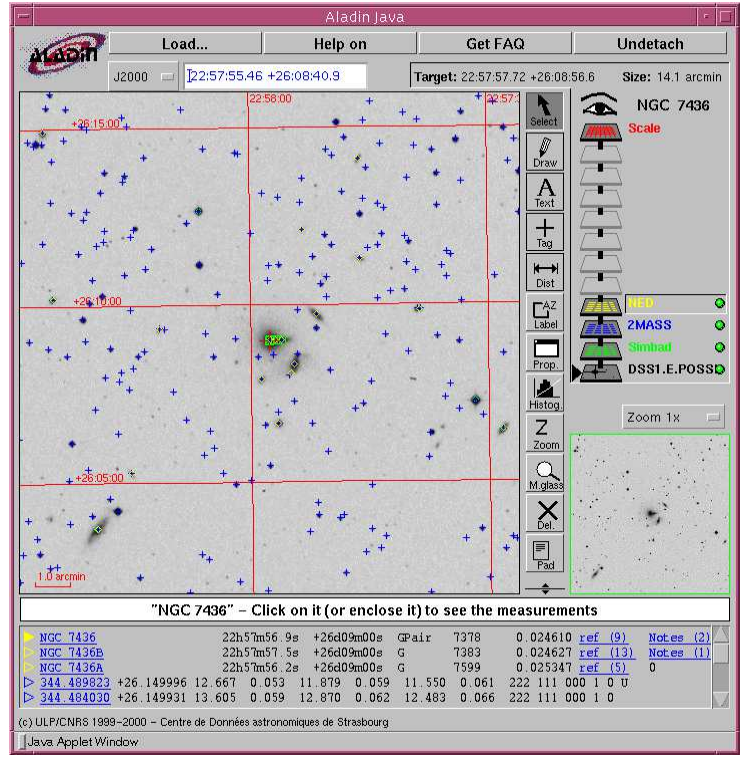,height=10cm}
   \end{tabular}
   \end{center}
   \caption[Aladin]{ \label{ngc7436}
Example of {\sc Aladin} window, with an image centered
on NGC 7436, and objects from SIMBAD, NED and 2MASS marked by symbols.}
   \end{figure}

The system, proposed as a Java applet
(see Figure~\ref{ngc7436}),
allows the user to visualize digitized images of a part of the sky,
to superimpose entries from the CDS astronomical catalogues and tables,
and to interactively access related data and information
from the {\sc Simbad}, NED, VizieR, or other archives for all known 
objects in the field.

{\sc Aladin} is particularly useful for 
multi-spectral cross-identifications of
astronomical sources, observation preparation and
quality control of new data sets (by comparison with
standard catalogues covering the same region of sky).

Data integration is made possible through the use of the
GLU data dictionary, and of the ASU protocol.
Input and output formats are gradually migrated
to  the AstroRes XML format (section~\ref{astrores}), 
as soon as it is implemented
in the remote services.

\subsection{ISAIA}
\label{isaia}

ISAIA (Integrated System for Archival Information 
Access\footnote{http://heasarc.gsfc.nasa.gov/isaia/};
Hanisch 2000a\cite{isaia1}, 2000b\cite{isaia2})
is a project to build an interoperability layer (middleware)
for providing access to distributed space science 
data resources via common query protocols and metadata standards.

The key objective of the project is to
develop an interdisciplinary data location and integration 
service for space sciences. Building upon existing data
services and communications protocols, this service will 
allow users to transparently query a large variety of distributed
heterogeneous Web-based resources (catalogues, data, computational 
resources, bibliographic references, etc.)
from a single interface. The service will collect responses 
from various resources and integrate them in a seamless
fashion for display and manipulation by the user. 

Because the scope of ISAIA is intended to span the space 
sciences -- astrophysics, planetary science, solar physics, and
space physics -- it is necessary to find a way to standardize the
descriptions of data attributes that are needed in order to formulate
queries.  The ISAIA approach is based on the concept of \emph{profiles}.
Profiles map generic concepts and terms onto mission or dataset specific
attributes.  Users may make general queries across multiple disciplines
by using the generic terms of the highest level profile, or make more
specific queries within subdisciplines using terms from more detailed
subprofiles.

The profiles play three critical and interconnected roles:
\begin{enumerate}
\item They identify appropriate resources (catalogues, mission datasets,
bibliographic databases):  the \emph{resource profile}
\item They enable generic queries to be mapped unambiguously onto 
resource-specific queries:  the \emph{query profile}
\item They enable query responses to be tagged by content type and 
integrated into a common presentation format:  the \emph{response
profile}
\end{enumerate}
The resource, query, and response profiles are all aspects of a common
database of resource attributes.  Current plans call for these profiles to 
be expressed using XML (eXtensible Markup Language, an emerging standard
which allows embedding of logical markup tags within a document) and to be 
maintained as a distributed database using the CDS GLU facility.

The profile concept is critical to a distributed data service where one
cannot expect data providers to modify their internal systems or services
to accommodate some externally imposed standard.  The profiles act as a
thin, lightweight interface between the distributed service and the
existing specific services.  Ideally the service-specific profile
implementations are maintained in a fully distributed fashion, with
each data or service provider running a GLU daemon in which that site's
services are fully described and updated as necessary.  Static services 
or services with insufficient staff resources to maintain a local GLU
implementation can still be included, however, as long as their profiles
are included elsewhere in the distributed resource database.  The profile
concept is not unique to space science, but would apply equally well to 
any distributed data service in which a common user interface is desired
to locate information in related yet traditionally separate disciplines.

\section{Multi-wavelength cross-identification}

A key science driver for interoperability 
in the context of the Virtual Observatory is the ability
to perform multi-wavelength cross-identification at a massive
scale, using reference surveys, and the wealth of observational
data already collected in our disciplinary field.

\subsection{Scientific motivations}

One of the very first scientific motivations for the Virtual Observatory
is to develop new tools for an efficient multi-wavelength approach,
providing a global view of the complete energy distribution of
the astronomical objects.
 This approach should encompass, e.g., the ability to select and retrieve
objects according to the particular shape of their spectrum.

Because of the massive data volumes, it is not feasible for
remote users to download a significant fraction of these data.
Interoperable data-mining services should therefore be developed,
so that the user can seamlessly issue joint queries to multiple
distributed databases (see e.g. Lawrence 2001\cite{lawrence}).

\subsection{The example of X-ray and optical cross-identification}

Recent works have shown the power of cross-identifying X-ray
and optical catalogues (see Rutledge et al.\ 2000\cite{Xid}).
As an example of a result of such a general cross-identification
process, Guillout et al. (1999\cite{rastyc}) have reported the detection
of a late-type stellar population in the direction of the Gould Belt 
among stars found by
cross-correlating the ROSAT All-Sky Survey with the Tycho catalogue.

Among the many projects under way, we would like
to mention the ClassX project proposed by T. Mc Glynn.
Conceived as a prototype of the Virtual Observatory,
ClassX ('Classifying the High Energy Universe') has been recently
approved by NASA.
The goal is to build an automated classifier for X-ray sources 
and to use it to try to 
distinguish the physical classes of all known X-ray objects. 
Massive datasets from the HEASARC, ST-MAST, and Chandra archives 
along with information from VizieR, 2MASS, FIRST and other systems 
will all be needed within this effort.

\subsection{ESO/CDS data mining project}

The Centre de Donn\'ees astronomiques 
de Strasbourg (CDS) has been focusing in the recent years on new
developments for the cross-identification of objects from large surveys. 
SIMBAD provides the identification of objects published in
the literature, or from reference catalogues; 
the VizieR catalogue browser can be used to browse through the
survey result catalogues, and to compare them to other catalogues and 
published tables; the ALADIN sky atlas allows the user to overlay 
survey catalogues on images of the sky, together with SIMBAD, NED, 
and catalogues and published tables from VizieR
(Bonnarel et al.\ 2000b\cite{crossid}). 

ESO and CDS have been developing data-mining tools in order
to allow users to access and combine the
information stored in the forms of catalogues at CDS, 
and catalogues and data at ESO (prototype
`ESO-CDS data mining project'; Ortiz et al.\ 1999{\cite{dmf}). 
The idea is that remote users can either submit their own data ``tables'' 
(as ASCII files) for comparison,  or extract information
from either ESO or CDS to cross correlate by position in the sky 
or by any of the parameters provided by the data
catalogues. 
An important point has been the development of knowledge structures 
with the purpose of facilitating the
description of the data to provide highly flexible data-mining options. 

Two knowledge detection structures were developed: one for 
\emph{astronomical object types}, and the other for \emph{column
content}. 
\begin{itemize}
\item The structure for object type resembles the structure used 
in SIMBAD, with a four level hierarchy; the
source to assign object types to the catalogues and tables 
is the standardized description file (ReadMe file)
developed at CDS and shared now by other data centers and journals. 

\item The structure related to column content was fully developed 
for this project. It contains 35 main categories and
has a four level hierarchy. Categories such as Photometry, 
Positions, Spectroscopy, Time and Physical Quantities
are amongst the most populated. 

A \emph{Unified Content Descriptor} (UCD) has been assigned to each 
of the columns in each of the tables accessed
with Vizier. For that, we developed an automatic UCD assignation procedure 
based on column name, column units, and
column description. 
\end{itemize}

The effort of unifying column description or object types,
although not directly visible to the end user, is critical
for the rigorous data organization and description which 
will pave the the way to new Virtual Observatory applications
in the domain of data mining and knowledge extraction.

\section{Final remarks: the Astrophysical Virtual Observatory}

Interoperability is one of the major challenges of the
Virtual Observatory.  The new paradigm of a federation
of astronomical archives (see Williams 2001\cite{williams}),
as opposed to a central master database, imposes novel approaches.
We want to build a \emph{semantic web}, 
where each service can feed data to another
service, and interactively perform multiple steps on the user's behalf,
so that only a small valuable piece of knowledge comes out and is
presented to the scientist.

An Interoperability Working Group,
chaired by Fran\c{c}oise Genova has been appointed by
the European {\sc Opticon} network
(Gilmore 2001\cite{opticon}), recognizing the essential scientific and
practical benefits resulting from cost-effective tools and standards
for improving access to, and exchange and usage of, data archives and
astronomical information services. 
The goals of this working group, to which international partners
are associated, is to study the practical tools required
for enhanced interoperability between distributed heterogeneous
services, providing scientists with transparent navigational
tools.

\end{document}